\documentclass[twocolumn,groupedaddress,amsmath,aps,longbibliography]{revtex4-1}
\usepackage{hyperref}
\usepackage{graphicx}
\usepackage{amsmath,amsfonts}
\usepackage{dcolumn}
\usepackage{bm}
\usepackage{color}
\usepackage{multirow}
\usepackage{float}
\usepackage{url}
\usepackage{physics}
\usepackage{subfigure}
\usepackage[utf8x]{inputenc}
\usepackage{soul}
\usepackage[normalem]{ulem}
\begin{document}

\title{The effect of Ni doping on the electronic structure and superconductivity in the noncentrosymmetric ThCoC$_2$}

\author{Gabriel Kuderowicz}
\email{gabriel.kuderowicz@fis.agh.edu.pl}                        
\address{Faculty of Physics and Applied Computer Science, AGH University of Science and Technology,
Aleja Mickiewicza 30, 30-059 Kraków, Poland}                        
\author{Bartlomiej Wiendlocha}
\email{wiendlocha@fis.agh.edu.pl}                        
\address{Faculty of Physics and Applied Computer Science, AGH University of Science and Technology,
Aleja Mickiewicza 30, 30-059 Kraków, Poland}                        

\date{\today}

\begin{abstract}
ThCoC$_2$ is a non-BCS, noncentrosymmetric superconductor with $T_c \simeq 2.5$~K, in which pairing mechanism was suggested to be either spin fluctuations or the electron-phonon coupling. 
Earlier experimental work revealed that $T_c$ can be greatly enhanced upon Ni doping in ThCo$_{1-x}$Ni$_x$C$_2$, up to 12~K for $x = 0.4$.
In this work, using the Korringa-Kohn-Rostoker method with the coherent potential approximation (KKR-CPA), the evolution of the electronic structure upon Ni doping was studied and the increase of the density of states at the Fermi level was found. 
The electron-phonon coupling constant $\lambda$ was calculated using the rigid muffin tin approximation (RMTA) and the increase in $\lambda(x)$ was found. This indirectly confirms the electron-phonon coupling scenario as the trend in $\lambda(x)$ is in qualitative agreement with the experiment when the electron-phonon pairing is assumed.
\end{abstract}

\maketitle

\section{Introduction}

Noncentrosymmetric superconductors attract constant attention as they may host superconductivity with a mixed spin singlet- spin-triplet character of the Cooper pairs~\cite{mixedstate1,bauer-book}. This is related to the lack of inversion symmetry and the presence of the antisymmetric spin-orbit coupling, which splits the bandstructure removing the spin degeneracy of the electronic bands.  
Examples of such superconductors are found among the strongly correlated heavy fermion compounds, like CePt$_3$Si, CeRhSi$_3$ or CeIrSi$_3$ \cite{Bauer2004,Kimura2007,Settai2008}, in which Cooper pairing is likely due to purely electronic interactions (spin fluctuations)~\cite{pairing-ceirsi3}, as well as among weakly correlated materials, where the electron-phonon coupling is more expected. In this latter group, one may find e.g. 
Li$_2$(Pd,Pt)$_3$B, Mg$_{10}$Ir$_{19}$B$_{16}$, or LaNiC$_2$ \cite{Togano2004,Badica2005,Yuan2006,Klimczuk2007,Lee1996,Wiendlocha2016}.

ThCoC$_2$ belongs to a large series of rare-earth carbides $R$NiC$_2$ and $R$CoC$_2$ which crystallize in a noncentrosymmetric, orthorhombic base-centered structure (Amm2, spacegroup no. 38). 
The $R$NiC$_2$ series was recently studied due to its interesting magnetic properties and charge density waves \cite{Kolincio2017,Kolincio2019}. 
ThCoC$_2$ was found to be a non-magnetic type-II superconductor below the critical temperature of T$_c \approx$ 2.5~K \cite{Grant2014}. 
Experimental studies revealed a non-BCS behavior of its superconducting phase.  
The temperature dependence of the electronic specific heat in the superconducting state 
strongly deviated from the exponential behavior, with the specific heat jump $\Delta C_e/\gamma T_c = 0.86$, much below the BCS value of 1.43 \cite{Grant2014}.
The normalized residual Sommerfeld coefficient as a function of a magnetic field showed a power-law dependence close to the square root relation, suggesting the presence of a nodal line in the superconducting gap \cite{Grant2017}.
The upper critical magnetic field {\it versus} temperature had positive curvature \cite{Grant2014}, indicating possibility of a multiband superconductivity.
Most recently, the magnetic field penetration depth was measured 
using the $\mu$-SR technique \cite{Bhattacharyya2019}. Its temperature dependence shows deviations from what is expected in the full-gap $s-$wave state and was fitted assuming a $d-$wave nodal superconducting gap. Moreover, spin fluctuations were suggested to be the pairing mechanism~\cite{Bhattacharyya2019}.

Effect of Ni doping was studied in Ref. \cite{Grant2017} and very interesting observations were made. Critical temperature in ThCo$_{1-x}$Ni$_x$C$_2$ was strongly increased, reaching $T_c = 12$~K for $x=0.4$. The alloy remains nonmagnetic for all Ni concentrations. Moreover, the non-BCS superconducting characteristics of pure ThCoC$_2$ appeared to be gradually suppressed with Ni substitution towards a more conventional fully gapped superconductor.

In our recent work \cite{Kuderowicz2021}, we have investigated the electronic structure, phonons, electron-phonon coupling, and superconductivity in ThCoC$_2$ with density functional theory. 
Large spin-orbit band splitting has been found, with an average value of $\Delta E_{SOC} \approx$ 150 meV, comparable to that in the triplet superconductors CePt$_3$Si and Li$_2$Pt$_3$B (see \cite{Smidman2017} and references therein). As the characteristic ratio, $E_r$ = {$\Delta E_{\rm SOC}/k_BT_c \simeq 700$} is large, 
in the limit of the strong spin-orbit interaction, the pairing inside the spin-split bands may require the odd parity of the gap with respect to the $\mathbf{k}\rightarrow -\mathbf{k}$~\cite{mixedstate2}, disturbing the $s$-wave symmetry.
At the same time, our calculations imply that the electron-phonon coupling constant $\lambda = 0.59$ is large enough to expect the electron-phonon interaction to be responsible for superconductivity in ThCoC$_2$.
To reveal how large are the deviations of the superconducting properties from the conventional isotropic $s-$wave state, we have calculated \cite{Kuderowicz2021} the thermodynamic properties of the superconducting phase by using the isotropic Eliashberg formalism \cite{Eliashberg1960}.
The temperature dependence of the London penetration depth, electronic heat capacity and magnetic critical field were computed. However, neither the BCS theory nor our more accurate Eliashberg solutions could reproduce the experimental results, strongly supporting the non-$s$-wave picture of superconductivity in ThCoC$_2$, but with the possibility of the electron-phonon coupling mechanism. 

The aim of the present work is to investigate the evolution of the electronic structure and electron-phonon coupling with Ni content in ThCo$_{1-x}$Ni$_x$C$_2$ to verify whether the electron-phonon coupling scenario is capable of explaining the substantial increase of $T_c$ determined experimentally in this alloy series.

\section{Calculations and results}

\begin{figure}[t]
	\centering
		\includegraphics[width=0.99\columnwidth]{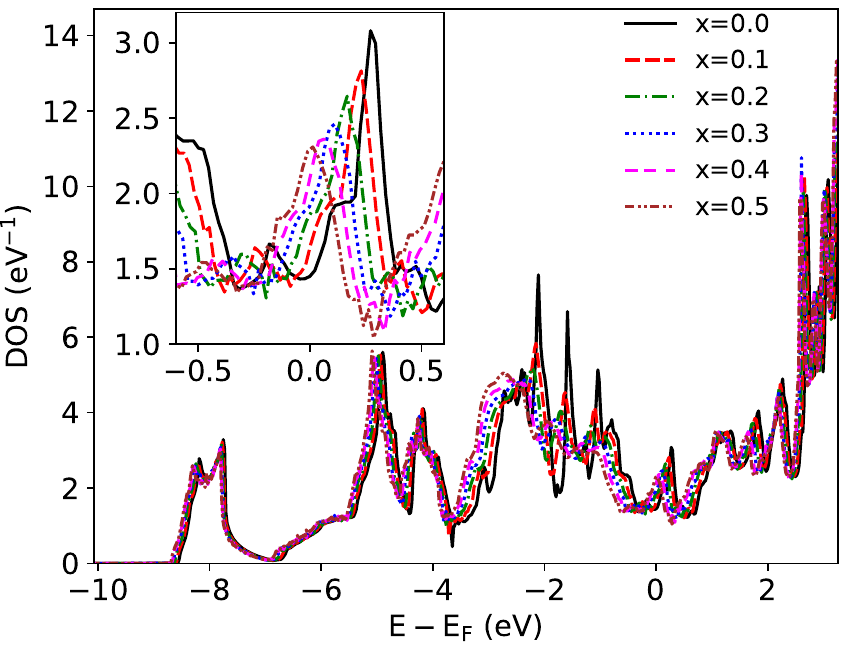}
	\caption{Density of states of ThCo$_{1-x}$Ni$_x$C$_2$. Inset shows zoomed region near the Fermi level.}
	\label{fig_1}
\end{figure}

The electronic structure was calculated using the Korringa-Kohn-Rostoker method with the coherent potential approximation (KKR-CPA) applied to account for the chemical disorder \cite{Bansil1999,stopa_2004}. 
Crystal potential of the spherical muffin-tin type was constructed using the local density approximation (LDA), with the Perdew-Wang parametrization~\cite{perdew_1992}, and in the semi-relativistic approach.
Angular momentum cut-off was set to $l_{max}$ = 3. 
Highly converged results were obtained using a {\bf k}-points mesh of 1728 points in the irreducible part of the Brillouin zone.
The Fermi level ($E_F$) was accurately determined from the generalized Lloyd formula \cite{kaprzyk_1990}. 
For each Ni concentration, the experimental lattice constants were used \cite{Grant2017} and atomic positions were kept fixed as in ThCoC$_2$~\cite{Kuderowicz2021}. Radii of muffin-tin spheres were adjusted to maximize structure packing.

To investigate superconductivity in the alloy series of ThCo$_{1-x}$Ni$_x$C$_2$ a 
simplified rigid muffin-tin approximation (RMTA) was used \cite{Gaspari1972}. It allows for separating the electronic and phonon contributions to $\lambda$, which is written as \cite{Wiendlocha2006a}:
\begin{equation}
    \lambda = \sum_i \frac{c_i\eta_i}{M_i \langle \omega_i^2 \rangle},
    \label{eq_lambda}
\end{equation}
where $\eta_i$ is the McMillan-Hopfield parameter of an atom $i$ with a mass $M_i$ and concentration $c_i$ in the unit cell, and $\langle \omega_i^2 \rangle = \int \omega F_i(\omega) d\omega / \int \omega^{-1} F_i(\omega) d\omega$ is an average phonon frequency calculated from partial phonon density of states $F_i(\omega)$. 
The McMillan-Hopfield parameter is defined as \cite{McMillan1968,Hopfield1969,Wiendlocha2006a}:
\begin{multline}
    \eta_i = \sum_l \frac{(2l+2)n_l^i(E_F)n_{l+1}^i(E_F)}{(2l+1)(2l+3)N(E_F)} \\ \times \left| \int_0^{R_{MT^i}} r^2 R_l^i(E_F,r) \frac{dV^i(r)}{dr} R_{l+1}^i(E_F,r) dr \right|^2,    
\end{multline}
where $l$ is the angular momentum number, $N(E_F)$ is the total density of states (DOS) at the Fermi level, $n_l^i(E_F)$ is a $l$-decomposed DOS, $V^i(r)$ is a spherical potential inside a muffin-tin of radius $R_{MT^i}$ and $R^i_{l+1}(E_F,r)$ is a normalized radial wavefunction. 
The discussion of the validity of the RMTA can be found in  \cite{Hopfield1969,Mazin1990,Wiendlocha2006b,Wiendlocha2008}.

\begin{figure}[t]
	\centering
		\includegraphics[width=0.99\columnwidth]{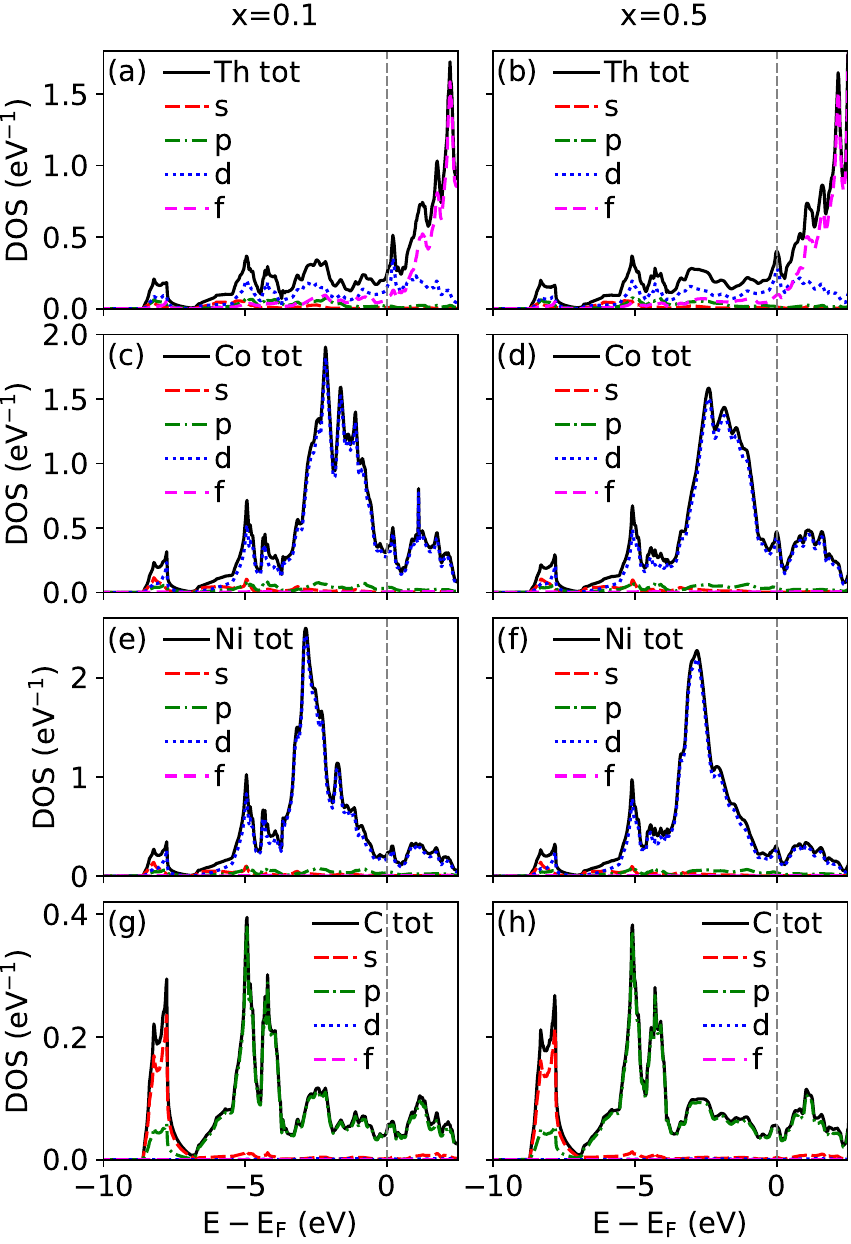}
	\caption{Projected density of states of ThCo$_{1-x}$Ni$_x$C$_2$ for x=0.1 (a,c,e,g) and x=0.5 (b,d,f,h).}
	\label{fig_2}
\end{figure}

In the present work, McMillan-Hopfield parameters were computed for each alloy concentration from the KKR-CPA bandstructure results, whereas the average atomic phonon frequency was computed from our previous results obtained for the pure ThCoC$_2$~\cite{Kuderowicz2021} using DFPT~\cite{Baroni2001}. 
For Ni, we assumed that $\langle \omega^2_{\rm Co} \rangle = \langle \omega^2_{\rm Ni} \rangle$, which can be justified because Co and Ni have very similar masses ($M_{Co}$ = 58.93~u and $M_{Ni}$ = 58.69~u).
Nonetheless, we are aware that this is a serious approximation since the real phonon spectrum will change with alloying (e.g, Debye temperature increases from $\theta_D =449$ K for $x=0$ to $\theta_D=584$ K for $x=0.4$). 

Computed DOS curves for $x \in [0.0; 0.5]$ with 0.1 step are shown in Fig.~\ref{fig_1}. 
For $x=0$, the overall shape of DOS is similar to that in Ref.~\cite{Kuderowicz2021}, however we observe a downward shift of the Fermi level towards the local minimum ($E_F$ is closer to the maximum in plane-wave calculations in \cite{Kuderowicz2021}).
This results in a diferent values of $N(E_F)$, 1.45~$\mathrm{(eV)^{-1}}$ compared to 2.14~$\mathrm{(eV)^{-1}}$.
The underestimation of $N(E_F)$ value for the starting system is most likely due to the spherical potential approximation used in the current work. The crystal of ThCoC$_2$ is a low-symmetry structure with only 4 symmetry operations, thus it is not surprising that the spherical potential is a crude approximation. Nevertheless, the qualitative discussion of the tendencies upon doping should remain valid.

\begin{figure}[t]
	\centering
		\includegraphics[width=0.99\columnwidth]{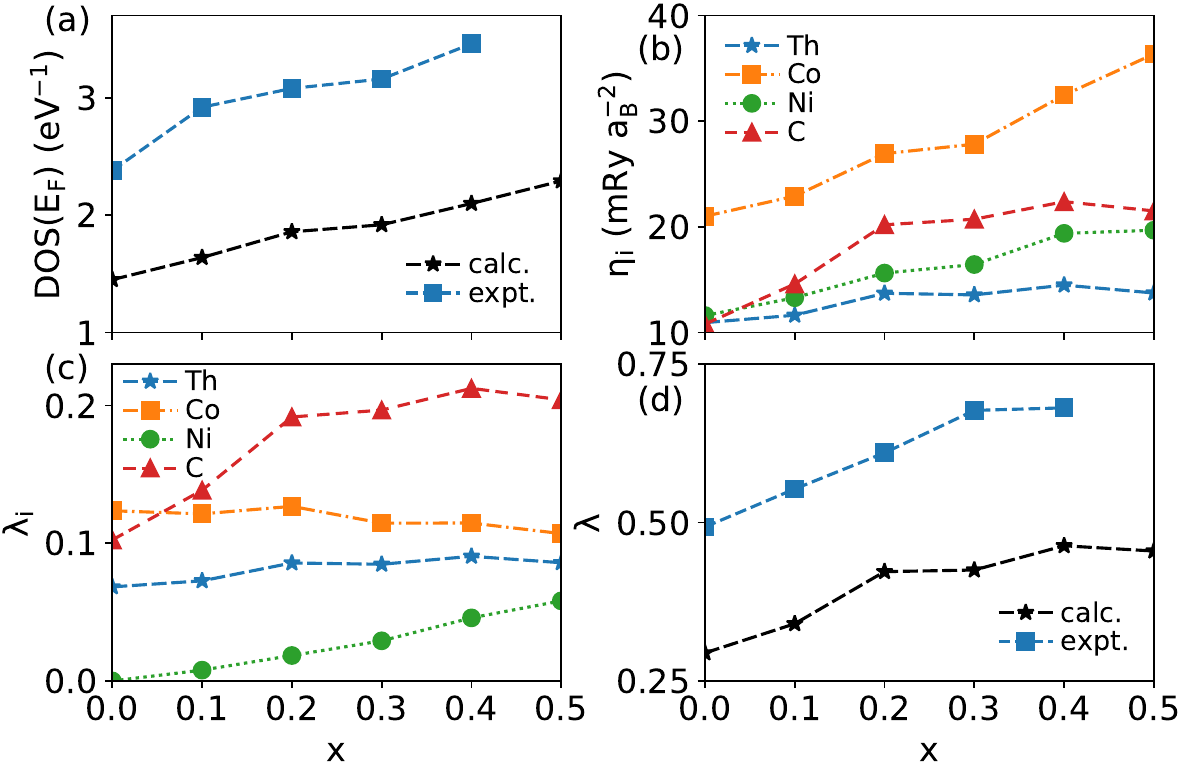}
	\caption{(a) Density of states at Fermi level of ThCo$_{1-x}$Ni$_x$C$_2$, (b) the McMillan-Hopfield parameters, (c) contributions to the electron-phonon coupling constant and (d) total electron-phonon coupling constant. Dashed lines are guides for the eye. {Points labeled as experimental are taken from Ref.~\cite{Grant2017}.}}
	\label{fig_4}
\end{figure}

\begin{table}[b]
  \caption{Density of states at Fermi level and McMillan-Hopfield parameters of ThCo$_{1-x}$Ni$_x$C$_2$.}
  \label{tab_1}
  \begin{tabular}{llllll}
    \hline
		\multirow{2}{*}{x} & $N(E_F)$ & $\eta_{Th}$ & $\eta_{Co}$ & $\eta_{Ni}$ & $\eta_{C}$\\
		 & $\mathrm{(eV^{-1})}$ & $\mathrm{(mRy\ a_B^{-2})}$ & $\mathrm{(mRy\ a_B^{-2})}$ & $\mathrm{(mRy\ a_B^{-2})}$ & $\mathrm{(mRy\ a_B^{-2})}$\\
    \hline
		0 & 1.45 & 10.930 & 21.001 & 11.605 & 10.791\\
        0.1 & 1.64 & 11.634 & 22.909 & 13.289 & 14.589\\
        0.2 & 1.86 & 13.709 & 26.934 & 15.625 & 20.196\\
        0.3 & 1.92 & 13.558 & 27.829 & 16.433 & 20.724\\
        0.4 & 2.10 & 14.485 & 32.478 & 19.384 & 22.385\\
        0.5 & 2.29 & 13.736 & 36.370 & 19.682 & 21.510\\
   \hline
  \end{tabular}
\end{table}

\begin{figure}[t]
    \centering
        \includegraphics[width=0.99\columnwidth]{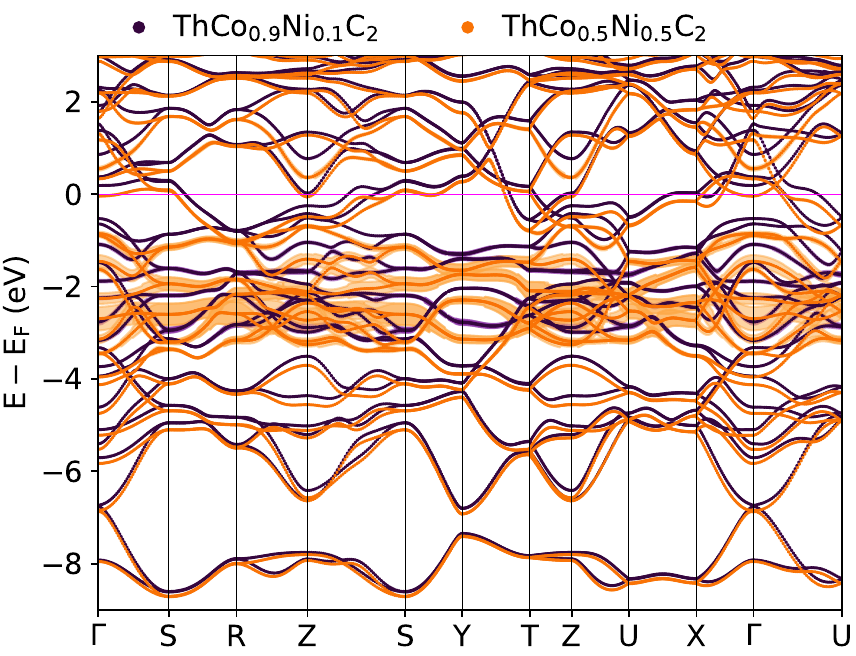}
    \caption{{Electronic bands of ThCo$_{1-x}$Ni$_x$C$_2$ for x=0.1 and x=0.5. Color shading represents band smearing (see text).}}
    \label{fig_band}
\end{figure}

Moving to the Ni substitution effect, as
Ni has one electron more than Co, this substitution moves the Fermi level with $x$ to higher energies in ThCo$_{1-x}$Ni$_x$C$_2$. In the inset of Fig.~\ref{fig_1} each DOS is plotted with respect to its Fermi level, and we can see that the local peak in DOS is smeared due to the alloying and shifted closer to $E_F$. 
Projected density of states is shown for two cases of $x = 0.1$ and $x=0.5$ in Fig. \ref{fig_2}. 
As in the undoped case~\cite{Kuderowicz2021}, we see a domination of the $d$-states of Th and Co (Ni) and $p-$states of C.
The increase of electron count with alloying almost linearly increases $N(E_F)$ value, as shown in Fig.~\ref{fig_4}(a). For $x=0.5$ $N(E_F)$ is increased by 50\%, see Table~\ref{tab_1}.
Nevertheless, in agreement with the experimental findings~\cite{Grant2017} the alloy remains nonmagnetic at the highest Ni concentration. 
Fig.~\ref{fig_4}(a) also shows the $N(E_F)$ trend extracted from the measurements of the electronic specific heat in Ref.~\cite{Grant2017}. The experimental $\lambda$ was calculated with the McMillan formula \cite{McMillan1968} using the Debye temperature acquired from specific heat measurements, $T_c$ and assuming Coulomb pseudopotential parameter $\mu^*$=0.13.
Our computed trend is in qualitative agreement with the experimental curve with an approximately constant systematic shift of $N(E_F)$ value.

{Electronic dispersion relations in the ThCo$_{1-x}$Ni$_x$C$_2$ alloy for the selected $x=0.1$ and $x=0.5$ cases are shown in Fig. \ref{fig_band}. They were computed
using the complex energy band technique~\cite{butler85}, where the energy eigenvalue has a real part, corresponding to the band center, and an imaginary part, related to the lifetime of the electronic state $\tau=\hbar/(2\ \mathfrak{Im}(E))$, which becomes finite due to scattering of electrons on impurities.
The imaginary part is visualized in Fig. \ref{fig_band} as the band shading, as it determines the strength of the band smearing effect induced by the atomic disorder.
As one can see, substituting Co with Ni shifts the bandstructure relative to the Fermi level to account for increasing number of electrons, with smaller changes in the shape of bands. For larger Ni content $E_F$ starts crossing bands in additional directions, e.g. near $\mathrm{\Gamma}$ and on the S-Y {\bf k}-path, which will lead to the appearance of additional Fermi surface pockets.
Band smearing effect is the strongest at 2-3 eV below the Fermi level, near the peak of the DOS of 3d states, most strongly affected by Ni alloying. As expected, band smearing increases with increasing the Ni concentration.
The above-mentioned changes in DOS and in the electronic dispersion relations upon alloying with Ni show that the  evolution of the electronic structure of ThCo$_{1-x}$Ni$_x$C$_2$ go beyond what one could expect based on the rigid band model.}

\begin{table}[t]
  \caption{Contributions and total electron-phonon coupling constant of ThCo$_{1-x}$Ni$_x$C$_2$ compared with the experiment \cite{Grant2017}.}
  \label{tab_2}
  \begin{tabular}{lllllllll}
  \hline
		\multirow{2}{*}{x} & \multirow{2}{*}{$\lambda_{Th}$} & \multirow{2}{*}{$\lambda_{Co_{1-x}}$} & \multirow{2}{*}{$\lambda_{Ni_{x}}$} & \multirow{2}{*}{$\lambda_{C_{2}}$} & \multirow{2}{*}{$\lambda_{tot}$} & \multirow{2}{*}{$\lambda_{\rm expt.}$} & $T_c^{\rm expt.}$ & $\gamma_{\rm expt.}$\\
		 &  &  &  &  &  &  & (K) & $\scriptstyle\mathrm{\left( \frac{mJ}{molK^2} \right)}$\\
		 \hline
        0   & 0.068 & 0.124 & 0     & 0.101 & 0.293 & 0.493 & 2.55 & 8.38\\
        0.1 & 0.073 & 0.121 & 0.008 & 0.137 & 0.339 & 0.553 & 4.40 & 10.70\\
        0.2 & 0.086 & 0.127 & 0.018 & 0.189 & 0.420 & 0.610 & 7.60 & 11.70\\
        0.3 & 0.085 & 0.115 & 0.029 & 0.194 & 0.423 & 0.677 & 10.40 & 12.50\\
        0.4 & 0.090 & 0.115 & 0.046 & 0.210 & 0.461 & 0.681 & 11.90 & 13.70\\
        0.5 & 0.086 & 0.107 & 0.058 & 0.202 & 0.453 & - & - & -\\
        \hline
  \end{tabular}
\end{table}

Calculated McMillan-Hopfield parameters $\eta_i$ are collected in Tab. \ref{tab_1} and their evolution with $x$ upon Ni alloying is presented in Fig~\ref{fig_4}(b).
Electron-phonon coupling constant $\lambda$, calculated using Eq.(\ref{eq_lambda}) is presented in Table~\ref{tab_2} and Fig.~\ref{fig_4}(c,d).
For the starting case of $x = 0$, similarly to the $N(E_F)$ value, the calculated $\lambda = 0.29$ is underestimated, comparing to $\lambda = 0.49$ extracted from the experimental $T_c$ as explained above, as well as to the value of $\lambda = 0.59$ computed from DFPT in our earlier work ~\cite{Kuderowicz2021}. $\lambda$ was calculated with the following frequencies $\sqrt{\langle\omega_i^2\rangle}$: 2.86 THz (Th), 5.85 THz (Co), and 14.53 THz (C).
The underestimation comes likely from the underestimated $N(E_F)$, as well as from the overall tendency of RMTA to underestimate $\lambda$ especially in the case of non-transition-metal elements~\cite{Hopfield1969,Mazin1990,Wiendlocha2006b,Wiendlocha2008}. From DFPT calculations~\cite{Kuderowicz2021} we have concluded that all atoms similarly contribute to the electron-phonon coupling, and RMTA seems to underestimate especially the thorium contribution. 

When Ni is substituted to the compound, all $\eta_i$ are increasing, enhancing the total electron-phonon coupling constant $\lambda$.
Table \ref{tab_2} and Fig.~\ref{fig_4}(d) compare the computed $\lambda(x)$ trend with the experimental data (from~\cite{Grant2017}, based on $T_c$ and the McMillan formula). 
The qualitative agreement between the computed and experimental trend is worth emphasizing, and the computed line is systematically shifted below the experimental one with an approximately constant value of $\sim 0.2$.
Substantial increase of computed $\lambda$ shows the strong enhancement of the electron-phonon coupling upon Ni substitution, caused mainly by the modification of the electronic structure. 
This qualitatively explains the experimentally observed increase in $T_c$ under the assumption of the electron-phonon coupling mechanism of superconductivity. 

\section{Summary}
In summary, we have calculated the electronic structure and McMillan-Hopfield parameters of ThCo$_{1-x}$Ni$_x$C$_2$ series as a function of Ni concentration, $x$. 
Ni causes the electron doping effect, moving the Fermi level closer to the local maximum of the density of states, which increases the $N(E_F)$ value.
Using the rigid muffin-tin approximation, we have obtained the electron-phonon coupling constant $\lambda(x)$, which strongly increases with $x$, by 57\% for $x=0.4$.
All these results remain in a qualitative agreement with experiments under the assumption of the electron-phonon coupling mechanism in ThCo$_{1-x}$Ni$_x$C$_2$.
Although due to the usage of the RMTA approach, our values of $\lambda(x)$ are systematically underestimated when compared to the experimental estimates, 
the qualitative agreement has to be noticed and supports the electron-phonon superconductivity scenario for ThCoC$_2$ outlined in Ref.~\cite{Kuderowicz2021}.

\section*{Acknowledgements}
This work was supported by the National Science Centre (Poland), project No. 2017/26/E/ST3/00119 and partly by the PL-Grid infrastructure.



\bibliography{refs}

\end{document}